\documentclass{aastex}
\usepackage{spr-astr-addons}
\usepackage{url}

\RequirePackage{color}

\newcommand{\emaila}{akimkin@inasan.ru}

\begin{document}

\title{UV-controlled physical and chemical structure \\ of protoplanetary disks}
\slugcomment{Not to appear in Nonlearned J., 45.}
\shorttitle{Structure of protoplanetary disks}
\shortauthors{Akimkin et al.}

\author{Akimkin V.V.\altaffilmark{1}} \and \author{Pavlyuchenkov Ya.N.\altaffilmark{1}}
\and
\author{Vasyunin A.I.\altaffilmark{2}\altaffilmark{3}}
\and
\author{Wiebe D.S.\altaffilmark{1}}
\and
\author{Kirsanova M.S.\altaffilmark{1}}
\and
\author{Henning Th.\altaffilmark{2}}
\email{\emaila}

\altaffiltext{1}{Institute of Astronomy of the RAS, Moscow, Russia}
\altaffiltext{2}{Max Planck Institute for Astronomy, Heidelberg, Germany}
\altaffiltext{3}{Ohio State University, USA}

\begin{abstract}
We study details of the UV radiation transfer in a protoplanetary disk, paying attention to the influence of dust growth and sedimentation on the disk density and temperature. Also, we show how the dust evolution affects photoreaction rates of key molecules, like CN and CS.
\end{abstract}



\section{Introduction}

Ultraviolet (UV) radiation is an important factor in the physical and chemical evolution of protoplanetary disks. It heats up a disk atmosphere, affecting both its structure and an emergent spectral energy distribution, and also controls the molecular content of the upper disk. A UV part of the spectrum comes primarily not from the star itself but from the inner region of an accretion flow (accretion shock, accretion column, etc.). As a result, the shape of the UV continuum and UV emission lines vary from a star to a star. Unlike the black body (BB) part of the stellar spectrum, its near UV part may significantly depend on details of a particular source, which is confirmed by available UV observations of T~Tau stars \citep{ref_uv_ttau}.

In this paper, we consider possible effects of a UV irradiation on the structure of a protoplanetary disk, in terms of density, temperature, and molecular composition. This influence is related to disk heating by the central source and to the rates of photoreactions. As both factors depend on the overall distribution of the UV field in the disk, it is reasonable to expect that they are most influenced by the grain evolution (growth and sedimentation) which marks the very initial stage of planet formation.

Radiative transfer~(RT) modeling of a protoplanetary disk is still a challenging problem (due to high optical depths and large variations
of physical conditions), and special efforts are being made to develop appropriate methods (see \cite{pas2004} for benchmarking of RT
codes). A large number of protoplanetary disk models has been
created over the last two decades. Some of these models are focused
on detailed (2D/3D) treatment of RT in dust continuum
in order to describe the disk thermal structure and its spectral
energy distribution (e.g. \cite{wolf2003}, \cite{nm2005},
\cite{dd2004}). The goal of other models is to study in detail
the chemical and micro-physical structure of protoplanetary disks
(e.g. \cite{sem2006}, \cite{nomura2007}, \cite{woitke2009}). These
simulations provide physical quantities (ionization degree, separate
dust and gas temperatures) which control the dynamics of
protoplanetary disks. These simulations are also extremely important
to interpret existing and future observations of dust and molecular
emission.

Our goal is to develop a ``balanced'' model of a protoplanetary disk
which would be relatively simple, but at the same time powerful enough
to allow direct observational verification. The model will be
balanced in terms of complexity/reliability between RT
and micro-physics treatment. With this model we plan to study the
influence of different physical processes on the physical/chemical
structure and observational appearance of a protoplanetary disk. In the
current version of the model we restrict ourselves to 1D vertical
RT but with scattering and detailed frequency dependence
(unlike \cite{woitke2009} where 2D RT is adopted but with only a few
frequency channels). As shown in \cite{dzn2002}, good
frequency coverage is more important than angular ray coverage, that
supports our approximation.
%

\section{Disk Model}

To find mutually consistent density and temperature distributions in a disk one needs an iterative process. As a starting point, we use gas and dust distributions from \cite{Wiebe_BD}. With these distributions being set, we can solve a RT problem and find thermal structure of the disk, accounting for the external radiation (a central star with UV excess and diffuse galactic UV background). The disk temperature distribution is found by means of a two-stream model of the RT with detailed frequency grid ($\simeq$ 200 frequencies covering a range
from 100 nm to 1 mm).

We assume that gas and dust are thermally coupled, while the temperature is determined by the dust thermal emission, absorption and isotropic scattering. The radiation field is separated into two streams, going from the surface of the disk to the midplane and in the opposite direction. Such simplified consideration leads to the following RT equation in the $z$-direction:
\begin{equation}
\frac{p}{\chi_{\nu}(z)}\frac{\partial}{\partial z}\left[\frac{1}{\chi_{\nu}(z)}\frac{\partial J_{\nu}(z)}{\partial z}\right]=J_{\nu}(z)-S_{\nu}(z)
\end{equation}
where $J_{\nu} [ \,\mbox{erg}\cdot \mbox{cm}^{-2}\cdot \mbox{s}^{-1}\cdot \mbox{Hz}^{-1}]$ is the mean intensity, $\chi_{\nu} [\,\mbox{cm}^{-1}]$ is the extinction coefficient, $p$ is equal to $1/3$ or $1/4$ for the cases of Eddington and Schwarzschild-Shuster approximations, respectively, and $S_{\nu}$ is the source function:
\begin{equation}
S_{\nu}=\frac{\varkappa_{\nu}B_{\nu}(T)+\sigma_{\nu}J_{\nu}}{\varkappa_{\nu}+\sigma_{\nu}}. 
\end{equation}
Here $B_{\nu}(T(z))$ is the Planck function, $\varkappa_{\nu} [\,\mbox{cm}^{-1}]$ and $\sigma_{\nu} [\,\mbox{cm}^{-1}]$ are the absorption and scattering coefficients, $\varkappa_{\nu}+\sigma_{\nu}=\chi_{\nu}$.  The RT equation has to be solved simultaneously with the energy balance equation which is written under the assumption of radiative equilibrium:
\begin{equation}
\int\limits_0^\infty\varkappa_{\nu}\left[B_{\nu}(T)-J_{\nu}\right]d\,\nu=0.
\end{equation}
 The RT equation is solved using the analog of the Feautrier method with the triagonal (Thomas) algorithm for a hypermatrix. To carry out the iterations between mean intensity $J_{\nu}$ and the temperature $T$ we employ the linear approximation of the Planck function $B_{\nu}(T)$.  The current approximation of $T(z)$ allows to find the updated mean intensity distribution $J_{\nu}(z)$. The solution of the RT equation in $z$-direction is repeated for a number of radial distances $R$, which gives the complete temperature distribution $T(R,z)$. The dust opacities are calculated from the Mie theory (astrosilicate dust graines, \cite{WD2001}).

Once the thermal structure $T(R,z)$ of the disk is known, we find the gas density structure $\rho(z)$ at each $R$ by integrating the equation of vertical hydrostatic equilibrium
\begin{eqnarray}
 \frac{\partial P(R,z)}{\partial z}&=&-\rho(R,z)\frac{zGM_{\star}}{\left(R^2+z^2\right)^{3/2}}, \label{GS}\\
  P&=&\frac{kT(R,z)}{\mu m_p}\rho(R,z),
\end{eqnarray}
where $P$ is the gas pressure, $\mu$ is the mean molecular weight ($\mu=2.3$ for H$_2$ and He mixture), $m_{\rm p}$ is the proton mass. Iterations are used to obtain self-consistent density and thermal distributions. In order to completely solve the problem we need to specify a surface density $\Sigma(R)$ which was taken from \cite{Wiebe_BD}.

The code has been extensively tested, first, for cases allowing
analytical solutions. We considered a) an optically thin media with
the only heating source (radiation or accretion), b) a media with
arbitrary optical depth irradiated by blackbody radiation field, and
c) one-frequency case for a media with arbitrary optical depth,
accretion heating and incident flux. Second, our model has been
compared with the model by \cite{dzn2002}. In their model,
the direct stellar irradiation is considered to be the separate dust
heating source, and the solution of the RT equation is being sought
for the diffuse field. Our model produces colder upper layers, while
midplane temperature is somewhat higher. We interpret the difference
in the upper layer temperature for the two models as a possible
consequence of underestimated direct stellar irradiation in our
model. At the same time, higher midplane temperatures in our model
seem to be more realistic due to more accurate computation of dust
radiative heating.

We consider two dust models and four representations for the spectrum of the central object. In model A5 dust is assumed to be well mixed with gas, with the mass ratio of 0.01. In model GS (growth and sedimentation) the dust distribution differs significantly from that of gas because of grain growth and sedimentation \citep{Birnstiel, vas2011}. While the average grain size is 10$^{-5}$ cm in model A5, in the midplane it increases because of dust evolution up to $3\cdot10^{-5}$~cm at the disk periphery ($R\sim500$\,AU) and $10^{-4}$~cm in the inner disk region (we only consider locations with $R>1$\,AU). Large grains sediment to the disk midplane, which causes spatial variations in the dust-to-gas mass ratio. The ratio grows to few times $10^{-2}$ in the midplane and decreases to $10^{-7}$ in the disk atmosphere.

The stellar spectrum is assumed to be that of a black body with $T_*=4000$\,K. We add a UV excess to this spectrum, described by the scaled interstellar UV field \citep[JD;][]{draine78}, a smoothed BP Tau spectrum \citep[JB;][]{kl2003}, and a black body spectrum with temperature of 20000\,K (J20). In all cases excess UV continuum is added only for $\lambda < 4000$\,\AA. All added UV contributions are normalized to have the same mean intensity at $\lambda = 4000$\,\AA, as the star spectrum. Also, we consider a case without a UV excess (J4). Model designations are listed in Table~\ref{moddes}.

\begin{table}
\caption{Model designations.}
\label{moddes}
\begin{tabular}{ll}
\hline
\multicolumn{2}{c}{\bf Dust models}\\
\hline
GS & Growth and sedimentation \\
A5 & Unevolving dust properties with \\
   & an average size of $10^{-5}$ cm \\
\hline
\multicolumn{2}{c}{\bf Incident UV spectra}\\
\hline
J4 & Black body spectrum with $T_*=4000$\,K\\
J20 & Black body spectrum with $T_*=20000$\,K\\
JB & Scaled BP Tau spectum\\
JD & Scaled interstellar Draine UV field\\
\hline
\end{tabular}
\end{table}

\section{UV irradiation and disk structure}

The influence of grain growth and settling on the disk structure has been studied in the literature a number of times \citep[e.g.][]{dal2006,an2006}. The general conclusion of these studies is that dust growth causes lower temperatures in the upper disk. The stellar radiation is mostly absorbed by small grains, and their number diminishes due to coagulation. On the other hand, dust sedimentation makes the upper disk more transparent and hotter. The resultant temperature is thus defined by the net effect of dust evolution.

\begin{figure*}[!t]
\includegraphics[width=0.7\textwidth]{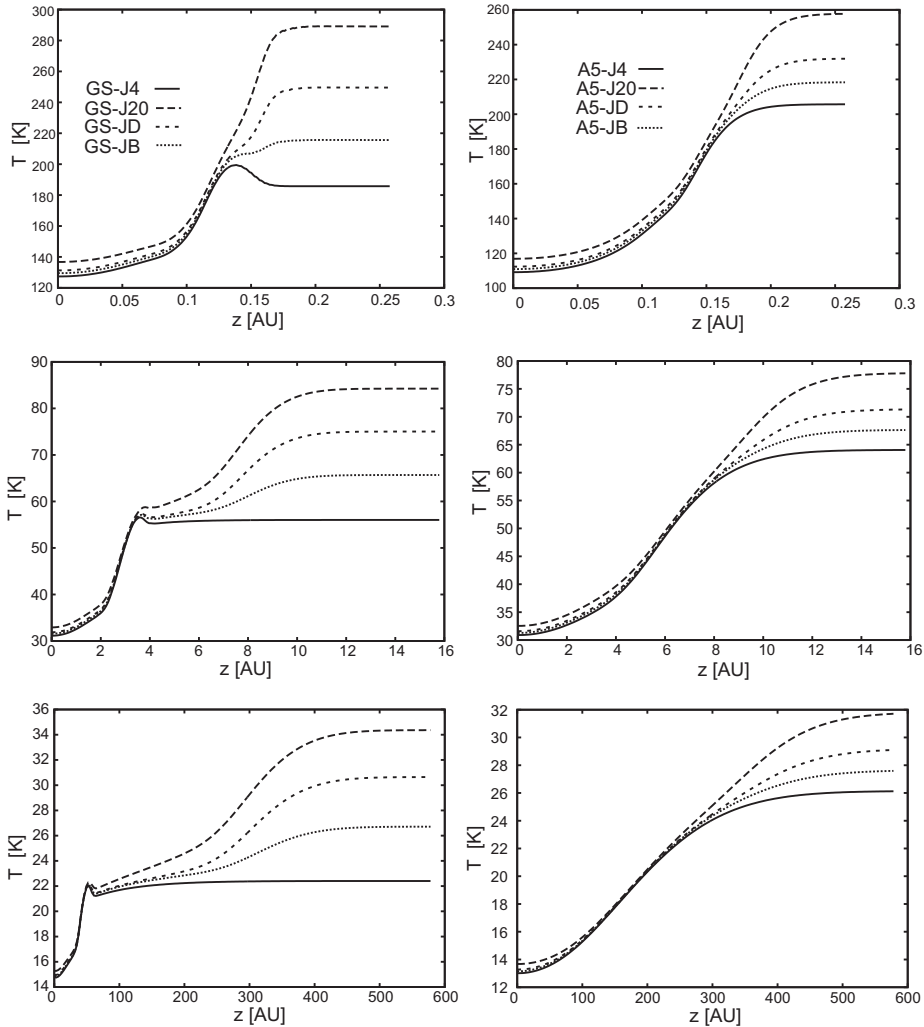}
\caption{Vertical temperature distributions at 1.4\,AU (top row), 33\,AU (middle row), and 508\,AU (bottom row) for evolved dust (left column) and pristine dust (right column).}
\label{fourcont}
\end{figure*}

Another factor which is important for disk heating is the UV continuum. In Figure~\ref{fourcont} we show vertical temperature profiles at distances of 1\,AU, 33\,AU, and 508\,AU from the central source for models GS (left column) and A5 (right column). For simplicity no iterations for the density structure has been performed for data on this figure. These profiles are quite different for the cases of pristine and evolved dust. In model A5, where dust has the standard ISM properties, vertical temperature profile only weakly depends on the shape of the UV spectrum of the central source, while in model GS the dependence is more significant. If no UV continuum is taken into account, the disk with evolved dust (GS-J4) is indeed somewhat colder than the disk with pristine dust (A5-J4). However, if there is UV continuum, the disk with evolved dust is hotter for considered cases. Temperature difference between various spectra is 120\,K in model GS at 1\,AU and less than 60\,K in model A5. A similar trend is observed at other radii as well. Deeper in the disk, temperature is less sensitive to the dust model. Quite expectedly, midplane temperatures are somewhat higher in the more transparent disk with evolved dust. It is seen that the temperature range is bracketed by J4 and J20 spectra so further we consider only these cases.

\begin{figure*}[!t]
\includegraphics[width=0.4\textwidth]{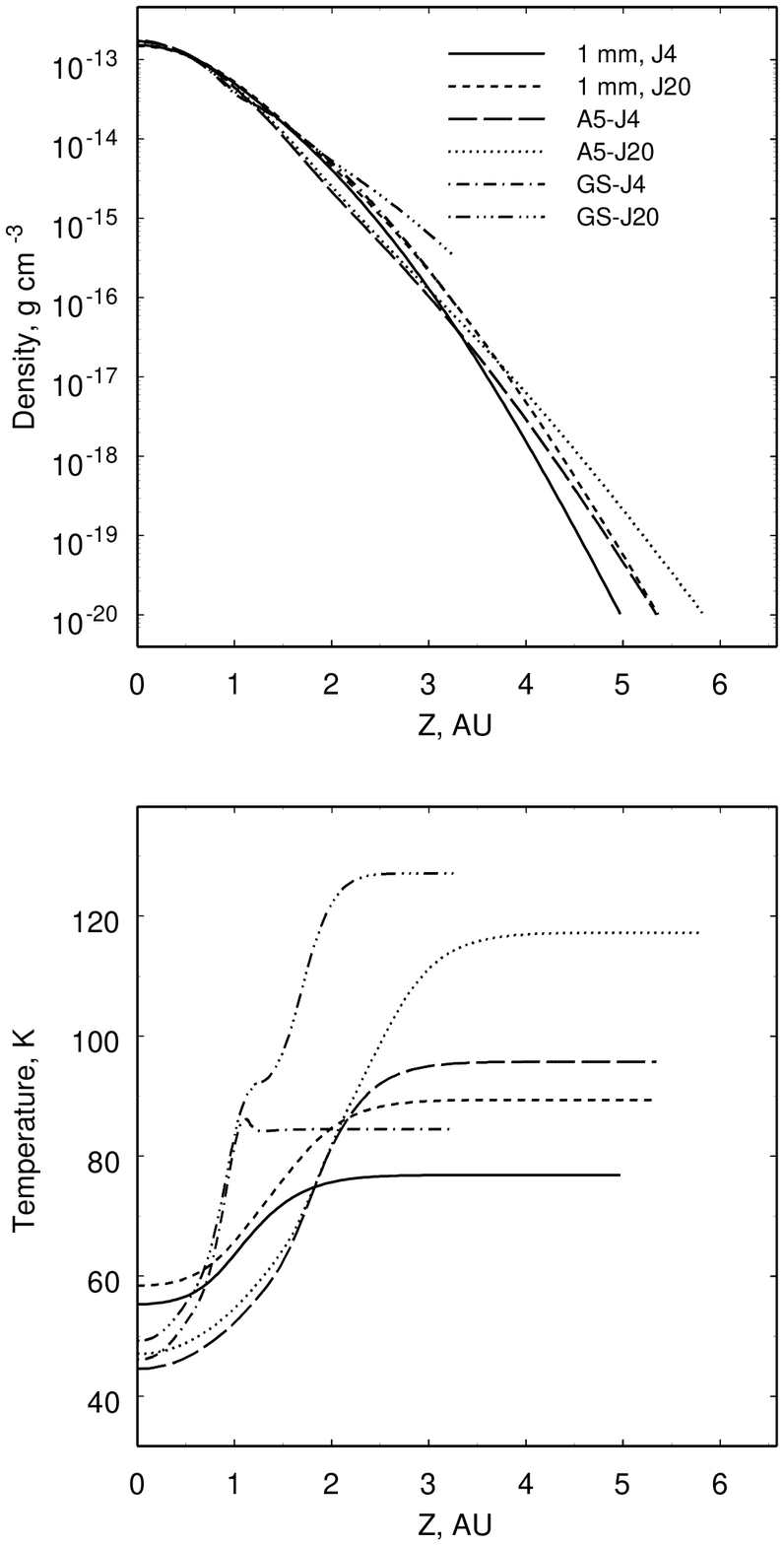}
\includegraphics[width=0.4\textwidth]{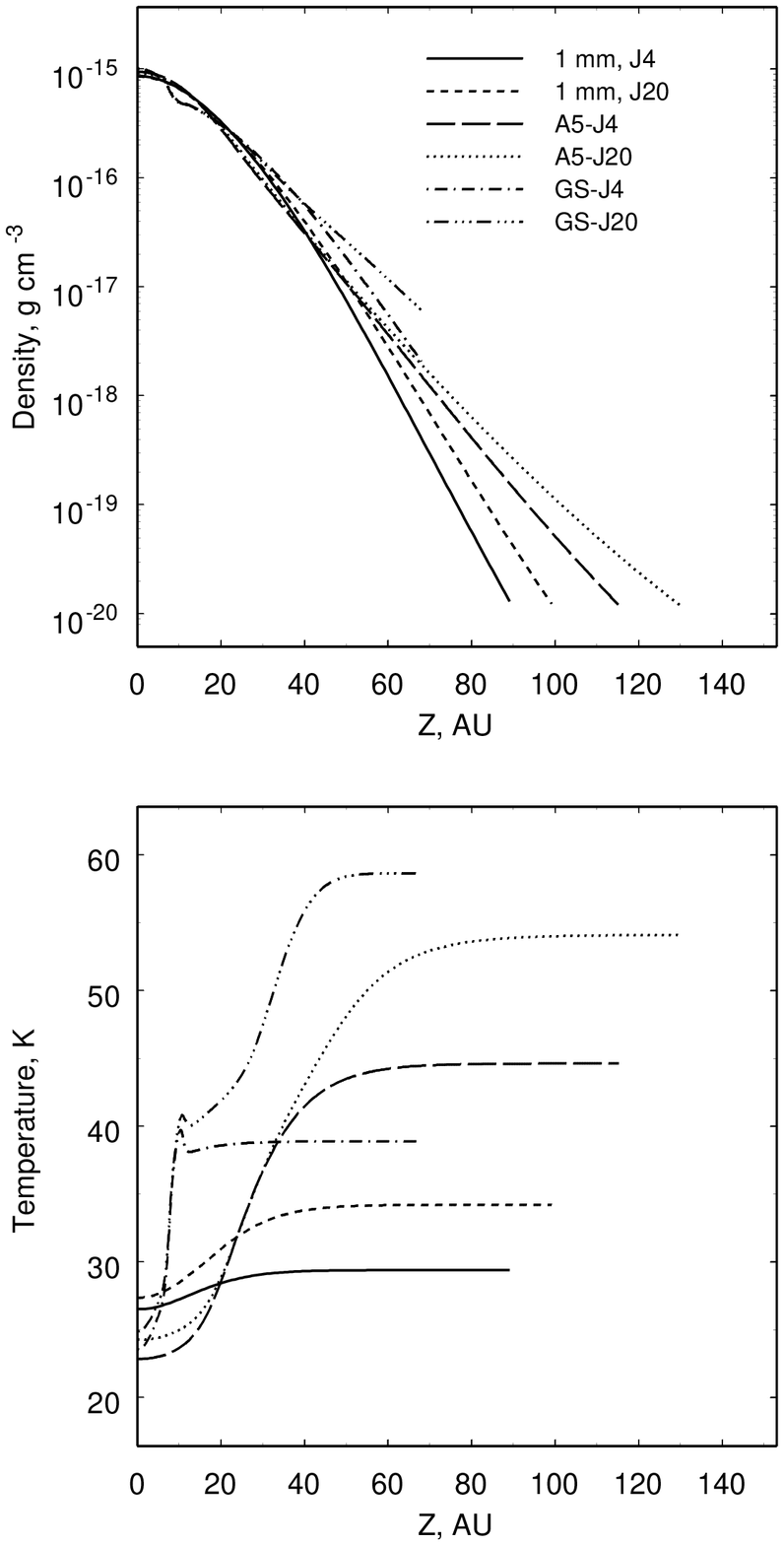}
\caption{Vertical density and temperature distributions at 10\,AU (left column) and 100\,AU (right column) for various dust models and stellar spectra.}
\label{hydro}
\end{figure*}

In Figure~\ref{hydro} we present results for the disk in hydrostatic equilibrium, illuminated by the central object with spectra J4 and J20. In addition to models GS and A5, we also show density and temperature structure for the model with the same size distribution as in model A5 but with the upper size limit of 1\,mm. It is seen that the upper limit of the grain size distribution does not change the density profile significantly, and corresponding curves in Figure~\ref{hydro} look very similar over the density drop of seven orders of magnitude. Model with evolved dust and  without UV continuum (GS-J4) follows this trend as well. However, the curve for model GS-J20 goes above all the other curves, indicating that the disk is puffed up in this model. It should be noted that curves for model GS end at lower heights than other curves. Dust density is very low at greater heights, and assumptions of our model break down. Obviously, a more detailed consideration is needed with separate treatment of dust and gas temperatures.

Temperature profiles are more diverse. Models with an upper grain size limit of 1\,mm are the coldest ones, in agreement with previous studies. Again, the importance of UV continuum is seen. For example, at 100\,AU it makes disk warmer by $\sim5$\,K in 1\,mm model, by $\sim10$\,K in model A5, and by $\sim20$\,K in model GS. This difference may not be strong enough to affect disk chemical composition, but it is of crucial importance for line transfer modeling. Note also that temperature gradient in model with evolved dust is much greater than in well-mixed models. This may cause differences not only in line intensities but also in the underlying disk molecular content.

\section{UV irradiation and disk chemistry}

\begin{figure*}[!t]
\includegraphics[width=0.6\textwidth]{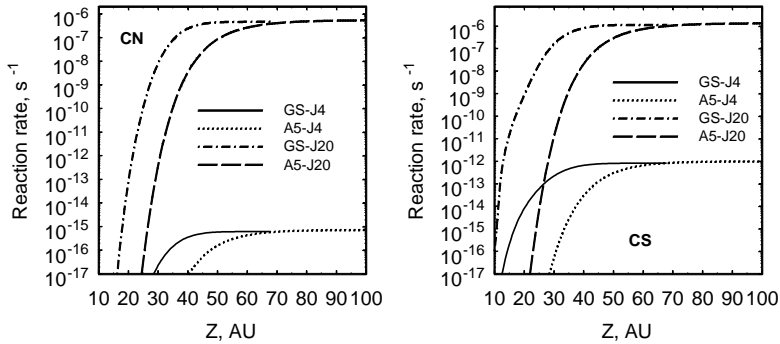}
\caption{Vertical profiles of CN and CS photodissociation rates at 100\,AU for various dust models and stellar spectra.}
\label{cncs}
\end{figure*}

UV irradiation of the disk coupled to the dust evolution also should have a profound effect on its molecular structure. To illustrate this, we use the disk models described above to compute photodissociation rates for CS and CN molecules (Figure~\ref{cncs}), for which detailed frequency-dependent cross-sections are available \citep{vds_fardi}. Again, we only show results for J4 and J20 spectra.

Obviously, photodissociation is much less effective in models without UV excess. Photodissociation rate of CS is lower by six orders of magnitude, while photodissociation rate of CN is lower by nine orders of magnitude. Also, these rates significantly depend on dust parameters, but this dependence is different for different molecules. For example, CN photodissociation rate reaches value of, say, $10^{-12}$~s$^{-1}$ at $z\approx30$\,AU in model A5-J20 and at $z\approx20$\,AU in model GS-J20. For CS photodissociation rate the corresponding height range is almost twice as large. Such a different behavior is related to wavelength dependence of reaction cross-sections. Molecules of CN are dissociated mostly by photons with $\lambda\sim1000$\,\AA, while CS molecules can be destroyed by photons with large $\lambda \sim1500$\,\AA. In model GS emission with longer wavelengths penetrates deeper into the disk due to dust settling, while emission with shorter wavelengths is effectively absorbed both in model A5 and in model GS.

\section{Conclusions}

In this paper we study the influence of UV continuum on the physical and chemical structure of protoplanetary disks. Disk parameters (thermal structure and chemical photoreaction rates) may sensitively depend both on the mere presence of UV continuum, and on its exact shape. Thus, the UV radiation affects disk observational appearance both in terms of continuum observations and molecular line observations. Influence of UV irradiation onto the disk structure gets stronger as dust particles grow bigger at the initial phase of planet formation. Thus, in order to interpret observations of a particular star+disk system its UV spectroscopy in the range of $1000-3000$\,\AA\ is greatly important, which makes future missions like WSO-UV highly desirable \citep{ass09,ass11}.
\newpage
\acknowledgments

This work is supported by the Federal Targeted Program ``Scientific and Educational Human Resources of Innovation-Driven Russia'' for 2009-2013. Ya. P. is supported by the grant MK-4713.2009.2. A. V. acknowledges the support of the National Science Foundation (US) for partial support of this work through grants AST-0702876 and GA10750-131847 to Eric Herbst.
\nopagebreak[4]
We are grateful to the anonymous referee for his/her comments and suggestions that helped us to improve the quality of our paper.
\nopagebreak[4]

\nopagebreak[4]

\end{document}